\documentclass[utf8]{frontiersSCNS} % for Science, Engineering and Humanities and Social Sciences articles
\usepackage{hyperref}
\usepackage{url,lineno,microtype}
\usepackage[onehalfspacing]{setspace}
\usepackage{amsmath}
\usepackage{graphicx}
\usepackage{float}
\usepackage{nameref}

\DeclareMathOperator*{\argmax}{arg\,max}

\def\keyFont{\fontsize{8}{11}\helveticabold}
\def\firstAuthorLast{Natekar {et~al.}} %use et al only if is more than 1 author
\def\Authors{Parth Natekar\,$^{1}$, Avinash Kori\,$^{1}$, and Ganapathy Krishnamurthi\,$^{1,*}$}
\def\Address{$^{1}$ Department of Engineering Design, Indian Institute of Technology Madras, Chennai}
\def\corrAuthor{Ganapathy Krishnamurthi}
\def\corrEmail{gankrish@iitm.ac.in}

\begin{document}
\onecolumn
\firstpage{1}

\title[Demystifying Brain Tumor Segmentation Networks]{Demystifying Brain Tumor Segmentation Networks: Interpretability and Uncertainty Analysis} 

\author[\firstAuthorLast ]{\Authors} %This field will be automatically populated
\address{} %This field will be automatically populated
\correspondance{} %This field will be automatically populated

\extraAuth{}

\maketitle
\begin{abstract}

The accurate automatic segmentation of gliomas and its intra-tumoral structures is important not only for treatment planning but also for follow-up evaluations. Several methods based on 2D and 3D Deep Neural Networks (DNN) have been developed to segment brain tumors and to classify different categories of tumors from different MRI modalities. However, these networks are often black-box models and do not provide any evidence regarding the process they take to perform this task. Increasing transparency and interpretability of such deep learning techniques is necessary for the complete integration of such methods into medical practice.  In this paper, we explore various techniques to explain the functional organization of brain tumor segmentation models and to extract visualizations of internal concepts to understand how these networks achieve highly accurate tumor segmentations.  We use the BraTS 2018 dataset to train three different networks with standard architectures and outline similarities and differences in the process that these networks take to segment brain tumors. We show that brain tumor segmentation networks learn certain human-understandable disentangled concepts on a filter level. We also show that they take a top-down or hierarchical approach to localizing the different parts of the tumor. We then extract visualizations of some internal feature maps and also provide a measure of uncertainty with regards to the outputs of the models to give additional qualitative evidence about the predictions of these networks. We believe that the emergence of such human-understandable organization and concepts might aid in the acceptance and integration of such methods in medical diagnosis.

\tiny
 \keyFont{ \section{Keywords: Interpretability, CNN, Brain Tumor, Segmentation, Uncertainty, Activation Maps, Features} } 
\end{abstract}

\section{Introduction}

Deep learning algorithms have shown great practical success in various tasks involving image, text and speech data. As deep learning techniques start making autonomous decisions in areas like medicine and public policy, there is a need to explain the decisions of these models so that we can understand \textit{why} a particular decision was made \citep{molnar2018interpretable}.

In the field of medical imaging and diagnosis, deep learning has achieved human-like results on many problems \citep{kermany2018identifying}, \citep{esteva2017dermatologist}, \citep{weng2017can}. Interpreting the decisions of such models in the medical domain is especially important, where transparency and a clearer understanding of Artificial Intelligence are essential from a regulatory point of view and to make sure that medical professionals can trust the predictions of such algorithms.

Understanding the organization and knowledge extraction process of deep learning models is thus important. Deep neural networks often work in higher dimensional abstract concepts. Reducing these to a domain that human experts can understand is necessary - if a model represents the underlying data distribution in a manner that human beings can comprehend and a logical hierarchy of steps is observed, this would provide some backing for its predictions and would aid in its acceptance by medical professionals.

However, while there has been a wide range of research on Explainable AI in general \citep{doshi2017towards}, \citep{gilpin2018explaining}, it has not been properly explored in the context of deep learning for medical imaging. \citep{holzinger2017we} discuss the importance of interpretability in the medical domain and provide an overview of some of the techniques that could be used for explaining models which use the image, omics, and text data.

In this work, we attempt to extract explanations for models which accurately segment brain tumors, so that some evidence can be provided regarding the process they take and how they organize themselves internally. We first discuss what interpretability means with respect to brain tumor models. We then present the results of our experiments and discuss what these could imply for machine learning assisted tumor diagnosis.

\section{Interpretability in the context of Brain Tumor Segmentation Models}

Interpreting deep networks which accurately segment brain tumors is important from the perspectives of both transparency and functional understanding (by functional understanding, we mean understanding the role of each component or filter of the network and how these relate to each other). Providing glimpses into the internals of such a network to provide a \textit{trace of its inference steps} \citep{holzinger2017we} would go at least some way to elucidating exactly how the network makes its decisions, providing a measure of legitimacy.

There have been several methods explored for trying to look inside a deep neural network. Many of these focus on visual interpretability, i.e. trying to extract understandable visualizations from the inner layers of the network or understanding what the network looks at when giving a particular output \citep{zhang2018visual}.

For a brain tumor segmentation model, such methods might provide details on how information flows through the model and how the model is organized. For example, it might help in understanding how the model represents information regarding the brain and tumor regions internally, and how these representations change over layers. 
Meaningful visualizations of the internals of a network will not only help medical professionals in assessing the legitimacy of the predictions but also help deep learning researchers to debug and improve performance.

\begin{figure}[h]
    \centering
    \includegraphics[width=1.0\textwidth]{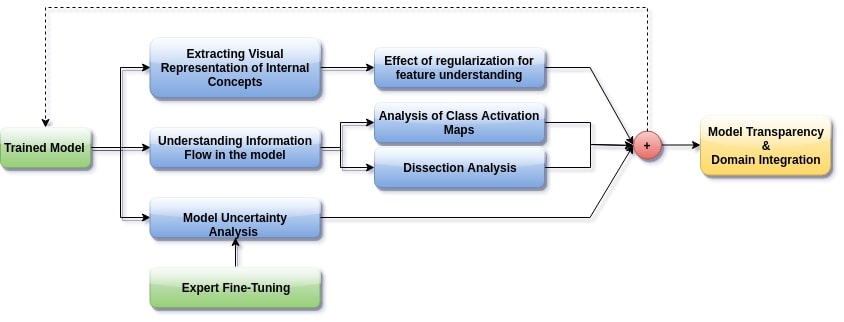}
    \caption{Proposed pipeline for interpreting brain tumor segmentation models to aid in increasing transparency. The dotted backward arrow shows the possiblity of using the inferences from such an experiment to enhance the training process of networks}
    \label{fig:pipeline}
\end{figure}
%For example, it might help in understanding what visual concepts a particular layer or filter learns, or how a network's understanding of such concepts changes over layers.

In this paper, we aim to apply visual interpretability and uncertainty estimation techniques on a set of models with different architectures to provide human-understandable visual interpretations of some of the concepts learned by different parts of a network and to understand more about the organization of these different networks.
We organize our paper into mainly three parts as described in Figure \ref{fig:pipeline}: (1) Understanding information organization in the model, (2) Extracting visual representations of internal concepts, and (3) Quantifying uncertainty in the outputs of the model. 
We implement our pipeline on three different 2D brain tumor segmentation models - a Unet model with a densenet121 encoder (Henceforth referred to as the DenseUnet) \citep{shaikh2017brain}, a Unet model with a ResNet encoder (ResUnet) \citep{kermi2018deep}, and a simple encoder-decoder network which has a similar architecture to the ResUnet but without skip or residual connections (SimUnet). All models were trained till convergence on the BraTS 2018 dataset (\cite{menze2014multimodal}, \cite{bakas2018identifying}, \cite{bakas2017advancing}, \cite{bakas2019segmentation}, \cite{bakas2020segmentation}). A held out validation set of 48 volumes (including both LGG and HGG volumes) was used for testing. Table \ref{tab:tab1} shows the performance of the three models on this test set.

\begin{table}[h]
\caption{Performance Metrics of our Networks. (WT: Whole Tumor, TC: Tumor Core, ET: Enhancing Tumor)}
\vspace{3mm}
\label{tab:tab1}
\centering
\begin{tabular}{|l|l|l|l|l|}
\hline
Model Type      & WT Dice & TC Dice & ET Dice \\ \hline
DenseUnet       & 0.830   & 0.760   & 0.685    \\ \hline
ResUnet         & 0.788   & 0.734   & 0.649   \\ \hline
SimUnet         & 0.743   & 0.693   & 0.523   \\ \hline

\end{tabular}
\end{table}

Our models are not meant to achieve state of the art performance. Instead, we aim to demonstrate our methods on a set of models with different structures commonly used for brain tumor segmentation and compare them to better understand the process they take to segment the tumors. In this primary study, we do not use 3D models, since the visualization and analysis of interpretability related metrics is simpler for 2D models. Also, it is not clear how some of our results would scale to 3D models and whether it would be possible to visualize these. For example, disentangled concepts observed by performing network dissection might not be meaningful when visualized slice wise and would have to be visualized in 3D. This and the related analysis poses an additional layer of difficulty. 

We now give a brief introduction of each interpretability techniques in our pipeline. \textit{Network Dissection} aims to quantify to what extent internal information representation in CNNs is human interpretable. This is important to understand what concepts the CNN is learning on a filter level, and whether these correspond with human level concepts. \textit{Grad-CAM} allows us to see how the spatial attention of the network changes over layers, i.e. what each layer of the network looks at in a specific input image. This is done by finding the importance of each neuron in the network by taking the gradient of the output with respect to that neuron. In \textit{feature visualization}, we find the input image which maximally activates a particular filter, by randomly initializing an input image and optimizing this for a fixed number of iterations, referred to as \textit{activation maximization}. Such an optimized image is assumed to be a good first order representation of the filter, which might allow us to to understand how a neural network 'sees'. \textit{Test-time dropout} is a computationally efficient method of approximate Bayesian Inference on a CNN to quantify uncertainty in the outputs of the model.

In the following sections, each element of the proposed pipeline is implemented and its results and implications are discussed.

\section{Understanding information organization in the model}
\label{section2}

\subsection{Network Dissection}
\label{dissection}

Deep neural networks may be learning explicit disentangled concepts from the underlying data distribution. For example, \citep{zhou2014object} show that object detectors emerge in networks trained for scene classification. To study whether filters in brain tumor segmentation networks learn such disentangled concepts, and to quantify such functional disentanglement (i.e. to quantify to what extent individual filters learn individual concepts), we implement the Network Dissection \citep{bau2017network} pipeline, allowing us to determine the function of individual filters in the network.

In-Network Dissection, the activation map of an internal filter for every input image is obtained. Then the distribution $\alpha$ of the activation is formulated over the entire dataset. The obtained activation map is then resized to the dimensions of the original image and thresholded to get a concept mask. This concept mask might tell us which individual concept a particular filter learns when overlaid over the input image. \par

For example, in the context of brain-tumor segmentation, if the model is learning disentangled concepts, there might be separate filters learning to detect, say, the edema region, or the necrotic tumor region. The other possibility is that the network somehow spreads information in a form not understandable by humans - entangled and non-interpretable concepts.

\noindent Mathematically, Network Dissection is implemented by obtaining activation maps $\Phi_{k, l}$ of a filter $k$ in layer $l$, and then obtaining the pixel level distribution $\alpha$ of $\Phi_{k,l}$ over the entire dataset.

A threshold $T_{k, l}(x)$ is determined as the 0.01-quantile level of $\alpha_{k, l}(x)$, which means only 1.0\% of values in $\Phi_{k, l}(x)$ are greater than $T_{k, l}(x)$. (We choose the 0.01-quantile level since this gives the best results qualitatively (visually) and also quantitatively in terms of dice score for the concepts for which ground truths are available). The concept mask is obtained as
\begin{equation}
    M_{k, l}(x) = \Phi_{k, l}(x) \geq T_{k, l}(x)
\end{equation}
A channel is a detector for a particular concept if 

\begin{equation}
    IoU(M_{k, l}(x), gt) = \dfrac{|M_{k,l}(x) \cap gt|}{|M_{k,l}(x) \cup gt|} \geq c
\end{equation}
In this study, we only quantify explicit concepts like the core and enhancing tumor due to the availability of ground truths $gt$ and recognize detectors for other concepts by visual inspection. We post-process the obtained concept images to remove salt-and-pepper noise and keep only the largest activated continuous concept inside the brain region in the image. The IoU between the final concept image and the ground truth for explicit concepts is used to determine the quality of the concept. 

\begin{figure}
    \centering
    \includegraphics[width=1\textwidth]{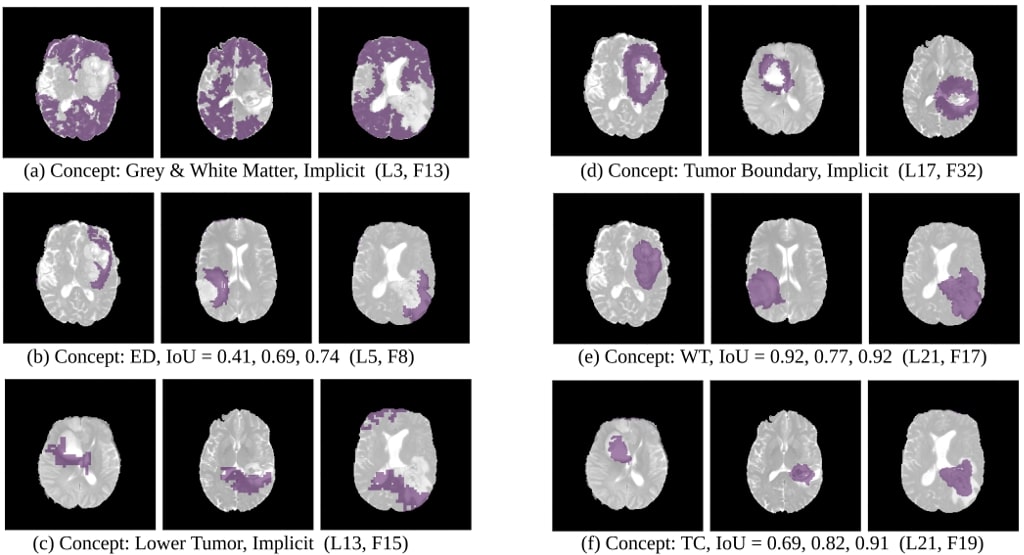}
    \caption{Disentangled concept mask $M$ learned by individual filters of the ResUnet overlaid over brain image. This includes explicit concepts for which ground truth labels are available as well as implicit concepts for which their are no labels. IoU scores are mentioned in the sub-captions for all 3 images.(L:Layer, WT: Whole Tumor, TC: Tumor Core, ED: Edema)}
    \label{fig:fig2a}
\end{figure}

\begin{figure}
    \centering
    \includegraphics[width=1\textwidth]{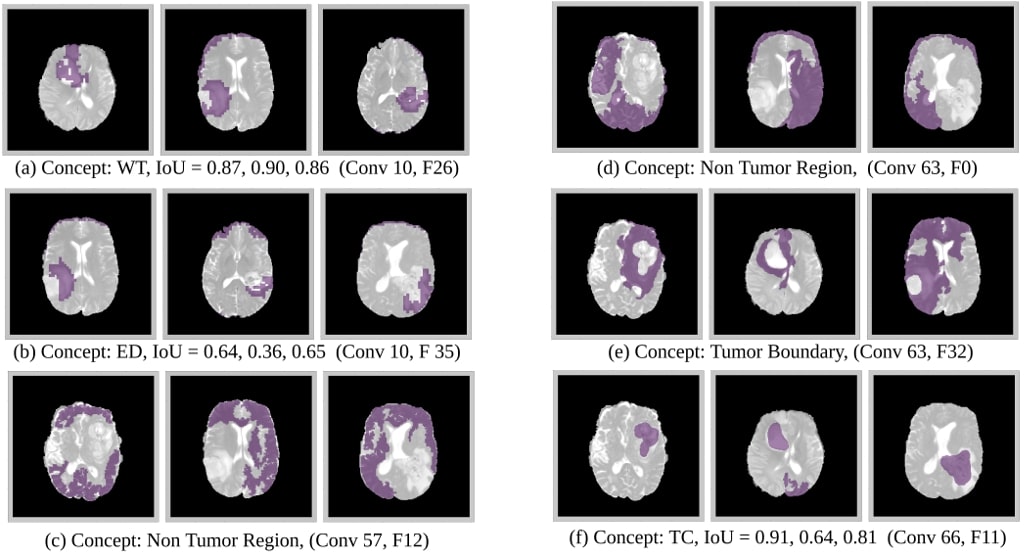}
    \caption{Disentangled concepts learned by filters of the DenseUnet. (L:Layer, WT: Whole Tumor, TC: Tumor Core, ED: Edema)}
    \label{fig:fig2b}
\end{figure}

\begin{figure}
    \centering
    \includegraphics[width=1\textwidth]{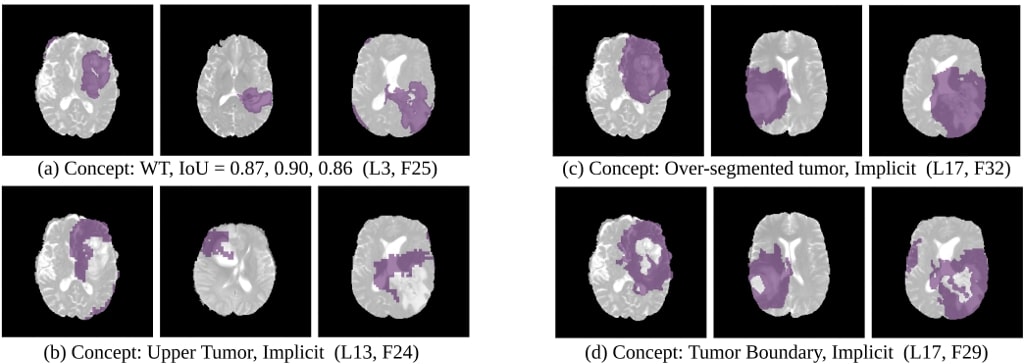}
    \caption{Disentangled concepts learned by filters of the SimUnet. (L:Layer, WT: Whole Tumor, TC: Tumor Core, ED: Edema)}
    \label{fig:fig2c}
\end{figure}

The results of this experiment, shown in Figures \ref{fig:fig2a}, \ref{fig:fig2b}, and \ref{fig:fig2c}, indicate that individual filters of brain-tumor segmentation networks learn explicit as well as implicit disentangled concepts. For example, Figure \ref{fig:fig2a}(e) shows a filter learning the concept \textit{whole tumor region} i.e. it specifically detects the whole tumor region for any image in the input distribution, the filter in \ref{fig:fig2a}(b) seems to be learning the \textit{edema region}, while \ref{fig:fig2a}(a) shows a filter learning the \textit{white and grey matter region}, an implicit concept which the network is not trained to learn. Similar behaviour is seen in all networks (Figures \ref{fig:fig2a}, \ref{fig:fig2b}, \ref{fig:fig2c}).This means that we can make attributions based on function to the network at a filter level - indicating a sort of functional specificity in the network i.e. individual filters might be specialized to learn separate concepts.

Neural Networks are inspired by neuroscientific principles. What does this functional specificity mean in this context? Debates are ongoing on whether specific visual and cognitive functions in the brain are segregated and the degree to which they are independent. \citep{american1998autonomy} discuss the presence of spatially distributed, parallel processing systems in the brain, each with its separate function. Neuroscientific studies have shown that the human brain has some regions that respond specifically to certain concepts, like the face fusiform area \citep{kanwisher2006fusiform} - indicating certain visual modularity. Studies based on transcranial magnetic stimulation of the brain also show separate areas of the visual cortex play a role in detecting concepts like faces, bodies, and objects \citep{pitcher2009triple}.

% \vspace{-5mm}
% \begin{figure}[H]
% \captionsetup[subfigure]{labelformat=empty, skip=2pt}
%     \centering
%     \subcaptionbox{(a) Concept: WT, IoU = 0.92, 0.77, 0.92 (L21,F17)} {\includegraphics[width=0.47\textwidth]{resnet_expli_21_17_whole.jpg}}\hfill
%     \subcaptionbox{(b) Concept: TC, IoU = 0.69, 0.82, 0.91 (L21,F19)} {\includegraphics[width=0.47\textwidth]{resnet_expli_21_19_TC.jpg}}\\
%     %vspace{-3mm}
%     \subcaptionbox{(c) Concept: ED, IoU = 0.41, 0.69, 0.74 (L5, F8)} {\includegraphics[width=0.47\textwidth]{resnet_expli_5_8_ed.jpg}}\hfill
%     \subcaptionbox{(d) Concept: tumor boundary, Implicit (L17,F31)} {\includegraphics[width=0.47\textwidth]{resnet_impli_17_31.jpg}}\\
%     %vspace{-3mm}
%     \subcaptionbox{(e) Concept: Lower Tumor, Implicit ([L13,F15)} {\includegraphics[width=0.47\textwidth]{resnet_impli_13_15.jpg}}\hfill
%     \subcaptionbox{(f) Concept: Grey \& White Matter, Implicit (L3,F13)} {\includegraphics[width=0.47\textwidth]{resnet_impli_3_13.jpg}}\\
%     %vspace{-3mm}
% \caption{Disentangled concept mask $M$ learned by individual filters of the ResUnet overlaid over brain image. This includes explicit concepts for which ground truth labels are available as well as implicit concepts for which their are no labels. IoU scores are mentioned in the sub-captions for all 3 images.(L:Layer, E:Encoding, B:Block, D:Decoding, WT: Whole Tumor, TC: Tumor Core, ED: Edima)}
    % \label{fig:fig2a}
% \end{figure}

%\vspace{-5mm}
% \begin{figure} [H]
% \captionsetup[subfigure]{labelformat=empty, skip=2pt}
% 	\centering
%     %\vspace{-3mm}
%     \subcaptionbox{(a) Concept: TC, IoU = 0.91, 0.64, 81 (DL10,F11)} {\includegraphics[width=0.49\textwidth]{dense_expli_10_11_TC.jpg}}\hfill
%     \subcaptionbox{(b) Concept: Non tumor Region, Implicit (DL8,F12)} {\includegraphics[width=0.49\textwidth]{dense_impli_8_12.jpg}}\\
%     %\vspace{-3mm}
%     \subcaptionbox{(c) Concept: WT, IoU = 0.80, 0.64, 0.70 (EL2,B1,F26)}{\includegraphics[width=0.49\textwidth]{dense_expli_dense_conv2_block1_1_conv_26_whole.jpg}}\hfill
%     \subcaptionbox{(d) Concept: Non tumor Region, Implicit (DL9,F1)}{\includegraphics[width=0.49\textwidth]{dense_impli_9_0.jpg}}\\
%     %\vspace{-3mm}
%     \subcaptionbox{(e) Concept: ED, IoU = 0.64, 0.36, 0.65 (EL2,B1,F35)}{\includegraphics[width=0.49\textwidth]{dense_expli_dense_conv2_block1_c1_35_edema.jpg}}\hfill
%     \subcaptionbox{(f) Concept: tumor Boundary, Implicit (L21,F17)}{\includegraphics[width=0.49\textwidth]{dense_impli_C9_32.jpg}}\\
    
% \caption{Disentangled concepts learned by filters of the DenseNet}
% \label{fig:fig2c}
% \end{figure}

% \begin{figure} [H]
% \captionsetup[subfigure]{labelformat=empty, skip=2pt}
% 	\centering
%     %\vspace{-3mm}
%     \subcaptionbox{(a) Concept: WT, IoU = 0.87, 0.90, 0.86 (L3,F25)}{\includegraphics[width=0.49\textwidth]{simnet_expli_3_25_whole.jpg}}\hfill
%     \subcaptionbox{(b) Concept: Over-segmented tumor, Implicit (L21,F17)}{\includegraphics[width=0.49\textwidth]{simnet_impli_17_32.jpg}}\\
%     %\vspace{-3mm}
%     %\subfloat[(c) Concept: tumor Core, IoU =  (L21,F32)]{\includegraphics[width=0.45\textwidth]{images/dissection2/SimNet/expli/21_32_TC.jpg}}\hfill
%     \subcaptionbox{(d) Concept: Upper tumor, Implicit (L21,F17)}{\includegraphics[width=0.49\textwidth]{simnet_impli_13_24.jpg}}\hfill
%     %\vspace{-3mm}
%     %\subfloat[(e) Concept: Whole tumor, IoU = 0.0 (L3,F7)]{\includegraphics[width=0.45\textwidth]{images/dissection2/SimNet/expli/3_7_whole.jpg}}\hfill
%     \subcaptionbox{(f) Concept: tumor Boundary, Implicit (L21,F17)}{\includegraphics[width=0.49\textwidth]{simnet_impli_17_29.jpg}}\\
%     %\vspace{4mm}
% \caption{Disentangled concepts learned by filters of the SimUnet}
% \label{fig:fig2b}
% \end{figure}

The emergence of concept detectors in our study indicates that brain-tumor segmentation networks might show a similar modularity. This indicates that there is some organization in the model similar to the process a human being might take to recognize a tumor, which might have an implications with regards to the credibility of these models in the medical domain, in the sense that they might be taking human-like, or at least human understandable, steps for inference.

The extracted disentangled concepts can also be used for providing contextual or anatomical information as feedback to the network. Though we do not explore this in this study, 3D concept maps obtained from networks can be fed back as multi-channel inputs to the network to help the network implicitly learn to identify anatomical regions like the gray and white matter, tumor boundary etc. for which no labels are provided, which might improve performance. This would be somewhat similar to the idea of feedback networks discussed by \cite{zamir2017feedback}, where an implicit taxonomy or hierarchy can be established during training as the network uses previously learned concepts to learn better representations and increase speed of learning.

\subsection{Gradient Weighted Class Activation Maps}

Understanding how spatial attention of a network over an input image develops might provide clues about the overall strategy the network uses to localize and segment an object. Gradient weighted Class Activation Maps (Grad-CAM) \citep{selvaraju2017grad} is one efficient technique that allows us to see the network’s attention over the input image. Grad-CAM provides the region of interest on an input image which has a maximum impact on predicting a specific class. \par

Segmentation is already a localization problem. However, our aim here is to see \textit{how attention changes over internal layers of the network}, to determine how spatial information flows in the model. To understand the attentions of each layer on an input image, we convert segmentation to a multi-label classification problem by considering class wise global average pooling on the final layer. The gradient of the final global average pooled value is considered for attention estimation in Grad-CAM. To understand the layer-wise feature map importance, Grad-CAM was applied to see the attention of every internal layer.

This mathematically amounts to finding neuron importance weights $\beta_{l, k}^c$ for each filter $k$ of a particular layer $l$ with respect to the global average pooled output segmentation for a particular channel $c$: 

% \begin{subequations} \label{eqn_2}
% \begin{align}
% y_c = \dfrac{1}{P}\sum_i\sum_j O_{ij}^c \\
% \alpha_k^c = \dfrac{1}{N}\sum_i\sum_j \dfrac{\delta y_c}{\delta A_{ij}^k} 
% \end{align}
% \end{subequations}
\begin{equation}
    y(c) = \dfrac{1}{P}\sum_i\sum_j \Phi^c (x) 
\end{equation}

\begin{equation}
    \beta_{l, k}^c = \dfrac{1}{N}\sum_{i} \sum_{j} \dfrac{\partial y(c)}{\partial A^{ij}_{l, k}(x)}
\end{equation}
\begin{equation}
    O_{GradCAM}(c) = ReLU\left (\sum_k \beta_{l, k}^c A_{l, k}(x) \right)  
\end{equation}

% \begin{subequations}
% \label{cam}
%     \begin{align}
%     y(c) = \dfrac{1}{P}\sum_i\sum_j \Phi^c (x)\\
%     \beta_{l, k}^c = \dfrac{1}{N}\sum_{i} \sum_{j} \dfrac{y(c)}{A^{ij}_{l, k}(x)}\\
%     O_{GradCAM}(c) = ReLU\left (\sum_k \beta_{l, k}^c A_{l, k}(x) \right)    
%     \end{align}
% \end{subequations}

Where, $P$ and $N$ are the number of pixels in the output segmentation map and the activation map of the relevant layer for channel $c$ respectively, $\Phi^c$ is the output segmentation map for class $c$ of network $\Phi$ , $y(c)$ describes the spatially pooled final segmentation map, $A_{l, k}(x)$ is the activation map for the $k^{th}$ filter of the $l^{th}$ layer, and  $O_{GradCAM}(c)$ represents an output map which is the result of $GradCAM$ for channel $c$.

% \begin{figure}[h]
% \captionsetup[subfigure]{labelformat=empty}
%     \centering
%   \subcaptionbox{(a) SimUnet}{\includegraphics[width=0.31\textwidth]{gradcam_simnet.jpg}}\hfill
%   \subcaptionbox{(b) ResUnet}{\includegraphics[width=0.31\textwidth]{gradcam_unet.jpg}}\hfill
%     \subcaptionbox{(c) DenseUnet}{\includegraphics[width=0.31\textwidth]{gradcam_dense.jpg}}\\
%     \subcaptionbox{}{\includegraphics[width=0.8\textwidth]{cb_label.jpg}}
%     \caption{This figure depicts the gradient based class activation maps obtained at selected intermediate layers of all the three networks in ascending order. (L:Layer, E:Encoding, B:Block, D:Decoding)}
%     \label{fig:fig3}
% \end{figure}

\begin{figure}[h]
\captionsetup[subfigure]{labelformat=empty}
    \centering
    \includegraphics[width=1.\textwidth]{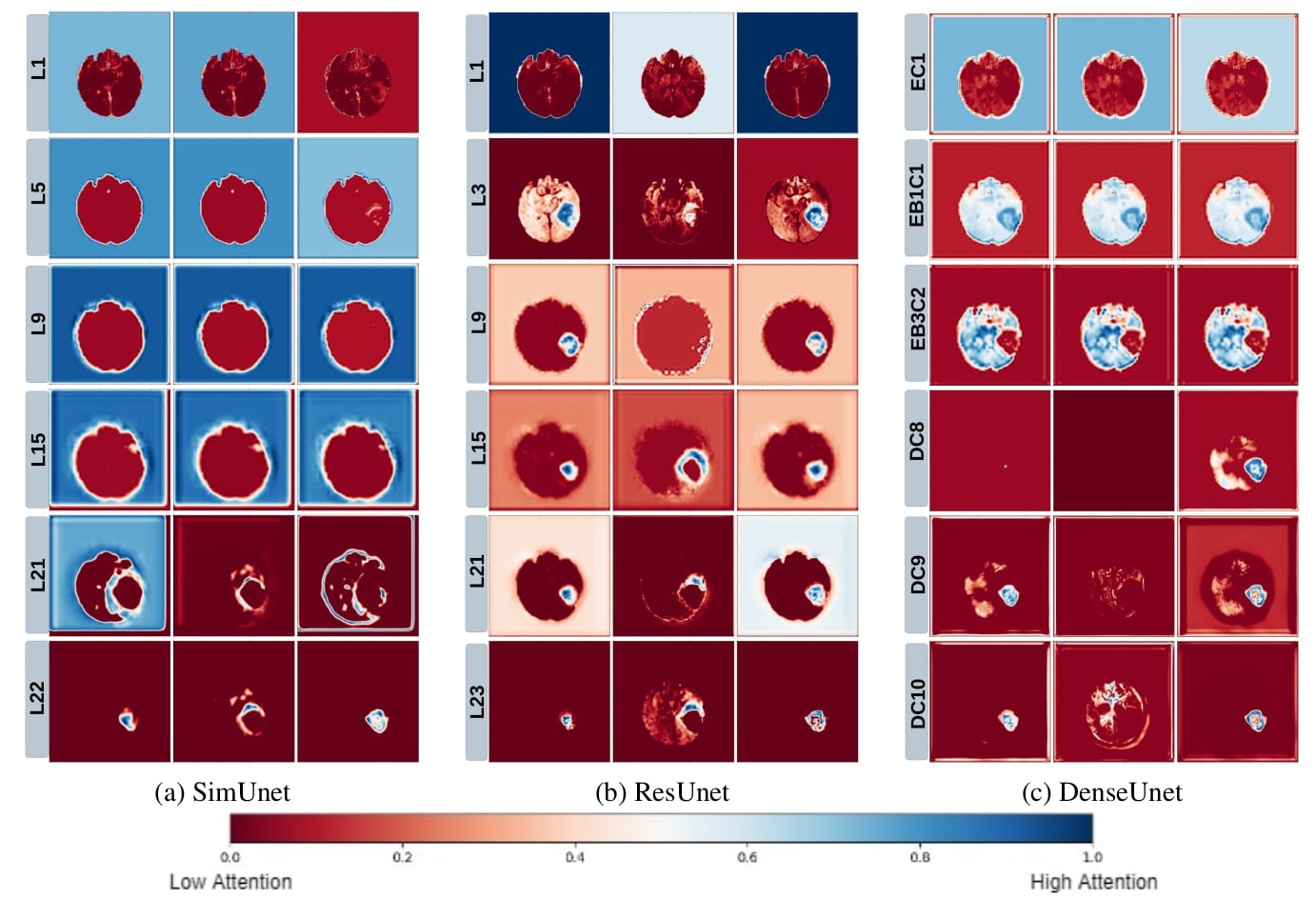}
    \caption{This figure depicts the gradient based class activation maps obtained at selected intermediate layers of all the three networks in ascending order. (L:Layer, E:Encoding, B:Block, D:Decoding)}
    \label{fig:fig3}
\end{figure}

We posit that model complexity and residual connections might have an impact on how early a model can localize the tumor region. For example, the DenseUnet and ResUnet localize the tumor region in the first few layers, while the SimUnet, which has no skip or residual connections, localizes the tumor region only in the final few layers (Figure \ref{fig:fig3}). This indicates that skip and residual connections help learn and propagate spatial information to the initial layers for faster localization. While previous literature indicates that skip connections allow upsampling layers to retain fine-grained information from downsampling layers \citep{jegou2017one}, \citep{drozdzal2016importance}, our results indicate that information might also be flowing in the other direction i.e. skip and residual connections help layers in the downsampling path to learn spatial information earlier.

\cite{drozdzal2016importance} also discuss that layers closer to the center of the model might be more difficult to train due to the vanishing gradient problem and that short skip or residual connections might alleviate this problem. Our results support this as well - middle layers of the SimUnet, which does not have residual or skip connections, seem to learn almost no spatial information compared to the other two networks (Figure \ref{fig:fig3}a).

Our results in Figure \ref{fig:fig3} also show that models take a largely top-down approach to localizing tumors - they first pay attention to the entire brain, then the general tumor region, and finally converge on the actual finer segmentation. For example, attention in all three models is initially in the background region. In the DenseUnet and ResUnet, attention quickly moves to the brain and whole tumor within the first few layers. Finer segmentations are done in the final few layers. The \textit{necrotic tumor} and \textit{enhancing tumor} are often separated only in the last few layers for all models, indicating that segregating these two regions might require a lesser number of parameters. 

This top-down nature is consistent with theories on visual perception in humans - the global-to-local nature of visual perception has been documented. \citep{navon1977forest} showed through experiments that larger features take precedence over smaller features, called the \textit{Global Precedence Effect}. While this effect has its caveats \citep{beaucousin2013global}, it is generally robust \citep{kimchi2015perception}. Brain tumor segmentation models seem to take a similar top-down approach, and we see in our experiments that such behavior becomes more explicit as model performance improves. 

While the results from the last two sections are not unexpected, they are not trivial either - the models do not need to learn disentangled concepts, especially implicit ones like the whole brain or the white matter region for which no explicit labels have been given, nor do they need to take a hierarchical approach to this problem. The fact that such human-understandable traces of inference can be extracted from brain tumor segmentation models is promising in terms of their acceptance in the medical domain.

\section{Extracting visual representations of internal concepts}

%Deep Neural Networks have the capacity to learn high-level features []. The increasing complexity of concepts learned by deep networks over different layers has been demonstrated on natural image data sets by visualizing features learned by the internal layers of the networks [].'
\subsection{Activation Maximization}

Visualizing the internal features (i.e. the representations of the internal filters obtained on activation maximization) of a network often provides clues as to the network's understanding of a particular output class. For example, visualizing features of networks trained on the ImageNet \citep{imagenet_cvpr09} dataset shows different filters maximally activated either by textures, shapes, objects or a combination of these \citep{olah2018the}. However, this technique has rarely been applied to segmentation models, especially in the medical domain. Extracting such internal features of a brain-tumor segmentation model might provide more information about the qualitative concepts that the network learns and how these concepts develop over layers. \par

We use the Activation Maximization \citep{erhan2009visualizing} technique to iteratively find input images that highly activate a particular filter. These images are assumed to be a good first-order representations of the filters. Mathematically, activation maximization can be seen as an optimization problem: 
\begin{equation} \label{eqn_1}
x^* = \argmax_x (\Phi_{k,l}(x) - R_\theta(x) - \lambda ||x||_2^2)
\end{equation}
Where, $x^*$ is the optimized pre-image, $\Phi_{k,l}(x)$ is the activation of the $k^{th}$ filter of the $l^{th}$ layer, and $R_\theta(x)$ are the set of regularizers.

In the case of brain-tumor segmentation, the optimized image is a 4 channel tensor. However, activation maximization often gives images with extreme pixel values or random repeating patterns that highly activate the filter but are not visually meaningful. In order to prevent this, we regularize our optimization to encourage robust images which show shapes and patterns that the network might be detecting.

\subsection{Regularization}

A number of regularizers have been proposed in the literature to improve the outputs of activation maximization. We use three regularization techniques to give robust human-understandable feature visualizations, apart from an L2 bound which is included in equation \ref{eqn_1}:

\subsubsection{Jitter}
In order to increase translational robustness of our visualizations, we implement Jitter \citep{inceptionism}. 
Mathematically, this involves padding the input image and optimizing a different image-sized window on each iteration. In practice, we also rotate the image slightly on each iteration. We find that this greatly helps in reducing high-frequency noise and helps in crisper visualizations. 

\subsubsection{Total Variation}

Total Variation (TV) regularization penalizes variation between adjacent pixels in an image while still maintaining the sharpness of edges \citep{strong2003edge}. We implement this regularizer to smooth our optimized images while still maintaining the edges.
The TV regularizer of an image I with (w, h, c) dimension is mathematically given as in equation \ref{tv}:

\begin{equation}
\label{tv}
\centering
R_{TV}(I) = \sum_{k=0}^c \sum_{u=0}^h \sum_{v=0}^w ([I(u, v+1, k) - I(u, v, k)] + [I(u+1, v, k) - I(u, v, k)])
\end{equation}

\subsubsection{Style Regularizer}
In order to obtain visualizations which are similar in style to the set of possible input images, we implement a style regularizer inspired from the work of \cite{li2017demystifying}. 
We encourage our optimization to move closer to the style of the original distribution by adding a similarity loss with a template image, which is the average image taken over the input data distribution. In style transfer, the gram matrix is usually used for this purpose. However, we implement a loss which minimizes the distance between the optimized and template image in a higher dimensional kernel space, as implemented in \cite{li2017demystifying}, which is computationally less intensive.

\noindent Mathematically, equation \ref{eqn_1} is modified to the following:

\begin{subequations}
\label{stylereg}
\centering
\begin{align}
x^* = \argmax_x(\Phi_{k,l}(x) - \zeta R_{TV}(x) + \gamma L(x,s)  - \lambda ||x||_2^2)\\
L(x, s) = \sum_i \sum_j (k(x_i,x_j) + k(s_i, s_j) - 2k(x_i, s_j))\\
k(x, y) = \exp(- \dfrac{||x-y||_2^2}{2\sigma^2})
\end{align}
\end{subequations}

 Where $L(x,s)$ it the style loss between the optimized pre-image and the template image $s$, $k(x, y)$ is the Gaussian kernel, $\Phi_{k,l}(x)$ is the filter for which activations need to be maximized,  $R_{TV}(x)$ is the Total Variation Loss, and $ ||x||_2^2$ is an upper bound on the optimized pre-image $x*$. Approximate values of the regularization coefficients are $\lambda \sim 10^{-4}$, $\gamma \sim 10^{-2}$, and $\zeta \sim 10^{-5}$. For jitter and rotation, the image is randomly shifted by $\sim$8 pixels, and rotated by $\sim$10 degrees.
 
 The effect of varying the hyperparameters for each of the regularizers is shown in Figure 13 in the supplementary material section. The effect of jitter is most pronounced - adding jitter by just 2-3 pixels helps reduce high frequency noise and clearly elucidate shapes in the image. Increasing total variation regularization increases smoothness while maintaining shapes and boundaries, reducing salt and pepper noise. Increasing style regularization brings the image closer to an elliptical shape similar to a brain. The effect of changing the regularization hyperparameters from a medical perspective in the context brain-tumor segmentation, however, is not clear and further studies would be required in this direction.

\begin{figure}[h]
\centering
\captionsetup[subfigure]{labelformat=empty}
    \includegraphics[width=1.\textwidth]{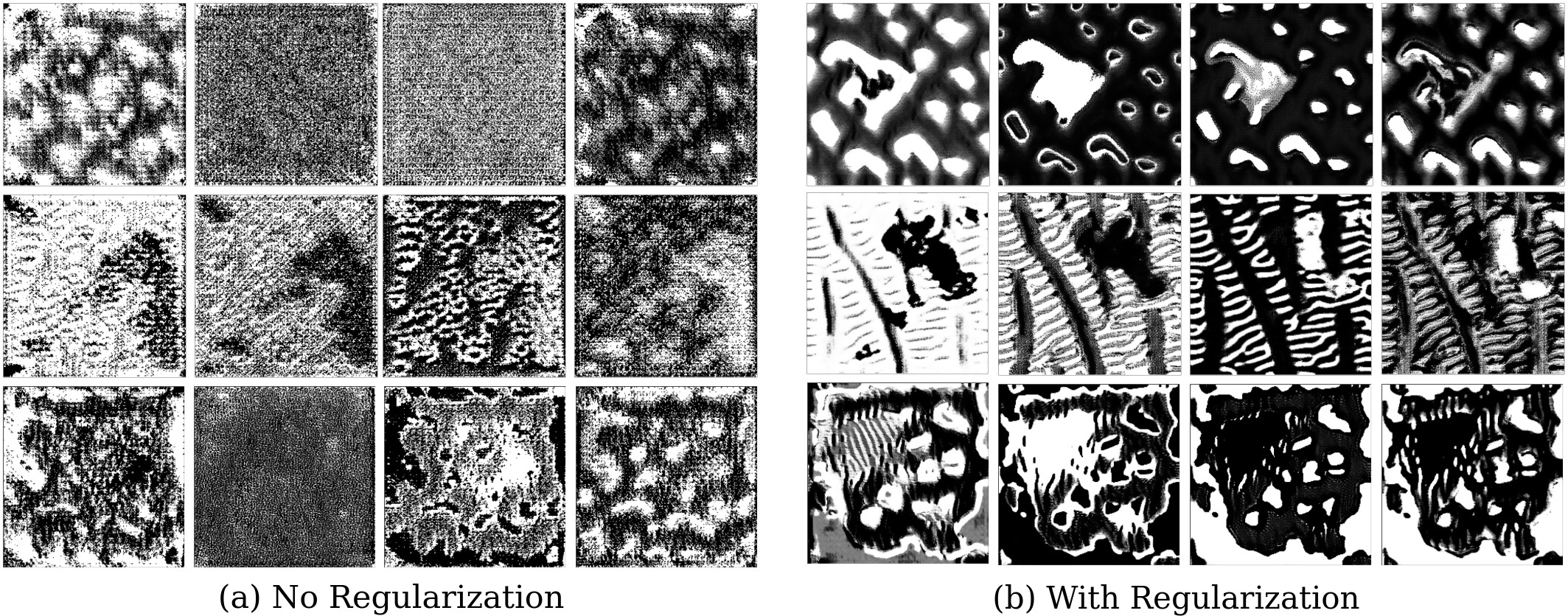}
    \caption{This figure depicts the effect of regularizers on visualized features of brain tumor segmentation models. The four columns on the left show the four channel feature map obtained on optimizing without regularization, while the columns on the right show the effect of adding regularizers}
    \label{fig:fig4}
\end{figure}

 We find that style constraining the images and making them more robust to transformations does help in extracting better feature visualizations qualitatively - optimized pre-images do show certain texture patterns and shapes. Figure \ref{fig:fig4} shows the results of such an experiment. The effect of regularizers is clear - not regularizing the image leads to random, repeating patterns with high-frequency noise. Constrained images show certain distinct shapes and patterns. It is still not clear, however, that these are faithful reflections of what the filter is actually detecting.

% \begin{figure}[h]
% \centering
% \captionsetup[subfigure]{labelformat=empty}
%   \subcaptionbox{No regularization}{\includegraphics[width=0.45\textwidth]{lucid_features_final_noreg.jpg}}\hfill
%     \subcaptionbox{With regularization}{\includegraphics[width=0.45\textwidth]{lucid_features_final_style.jpg}}
%     \caption{This figure depicts the effect of regularizers on visualized features of brain tumor segmentation models. The first column in both the subplots denotes the disentangled concept learnt by a specific feature map and next 4 columns are the four channeled input which maximizes the activation at that feature map}
%     \label{fig:fig4}
% \end{figure}

Not a lot of prior work has been done in this area in the context of medical imaging, and our results are useful in the sense that they show that constrained optimization generates such patterns and shapes as compared to noisy unregularized images, which has also been seen in the domain of natural images. In the natural image domain, the resulting pre-images, after regularization, have less high frequency noise and are more easily identifiable by humans. As discussed in the work of \cite{olah2017feature} and \cite{nguyen2016multifaceted}, jitter, L2 regularization, Total Variation, and regularization with mean images priors are shown to produce less noisy and more useful objects or patterns. In medical imaging, however, the resulting patterns and shapes are harder to understand and interpret.

In order to extract clinical meaning from these, a comprehensive evaluation of which regularizers generate medically relevant and useful images based on collaboration with medical professionals and radiologists would be required. This could provide a more complete understanding of what a brain tumor segmentation model actually detects qualitatively. However, this is out of scope of the current study. As we have mentioned in Section \ref{section7}, this will be explored in future work.

% We observe that while it is difficult to extract diagnostic meaning from the results of feature visualization, textures and patterns are visible on constraining the optimization to a more probable domain. 
% However, collaboration with radiologists and medical professionals in this context is required and could provide a complete understanding of what a brain tumor segmentation model actually detects qualitatively.

\section{Uncertainty}

Augmenting model predictions with uncertainty estimates are essential in the medical domain since unclear diagnostic cases are aplenty. In such a case, a machine learning model must provide medical professionals with information regarding what it is not sure about, so that more careful attention can be given here.
\cite{begoli2019need} discuss the need for uncertainty in machine-assisted medical decision making and the challenges that we might face in this context.

Uncertainty Quantification for deep learning methods in the medical domain has been explored before. \cite{leibig2017leveraging} show that uncertainties estimated using Bayesian dropout were more effective and more efficient for deep learning-based disease detection. \cite{yang2017quicksilver} use a Bayesian approach to quantify uncertainties in a deep learning-based image registration task. 

However, multiple kinds of uncertainties might exist in deep learning approaches - from data collection to model choice to parameter uncertainty, and not all of them are as useful or can be quantified as easily, as discussed below.

% \subsection{Epistemic Uncertainty}

Epistemic uncertainty captures uncertainty in the model parameters, that is, the uncertainty which results from us not being able to identify which kind of model generated the given data distribution.
Aleatoric uncertainty, on the other hand, captures noise inherent in the data generating process \citep{kendall2017uncertainties}. However, Aleatoric Uncertainty is not really useful in the context of this work - we are trying to explain and augment the decisions of the model itself, not the uncertainty in the distribution on which it is fit.

Epistemic uncertainty can, in theory, be determined using Bayesian Neural Networks. However, a more practical and computationally simple approach is to approximate this Bayesian inference by using dropout at test time.
We use test time dropout (TTD) as introduced in \citep{gal2016dropout} as an approximate variational inference.
Then, 

\begin{subequations}
\centering
\begin{align}
 p(y|x, w) \approx \dfrac{1}{T} \sum_{t=1}^t \Phi(x|{w^t})\\
 var_{epistemic}(p(y|x,w)) \approx \dfrac{1}{T} \sum_{t=1}^T \Phi(x|{w^t})^T \Phi(x|{w^t}) - \mathbf{E}(\Phi(x|{w^t}))^T \mathbf{E}(\Phi(x | {w^t}))
\end{align}
\label{eqn_unc}
\end{subequations}

\begin{figure}
    \centering
    \includegraphics[width=0.92\textwidth]{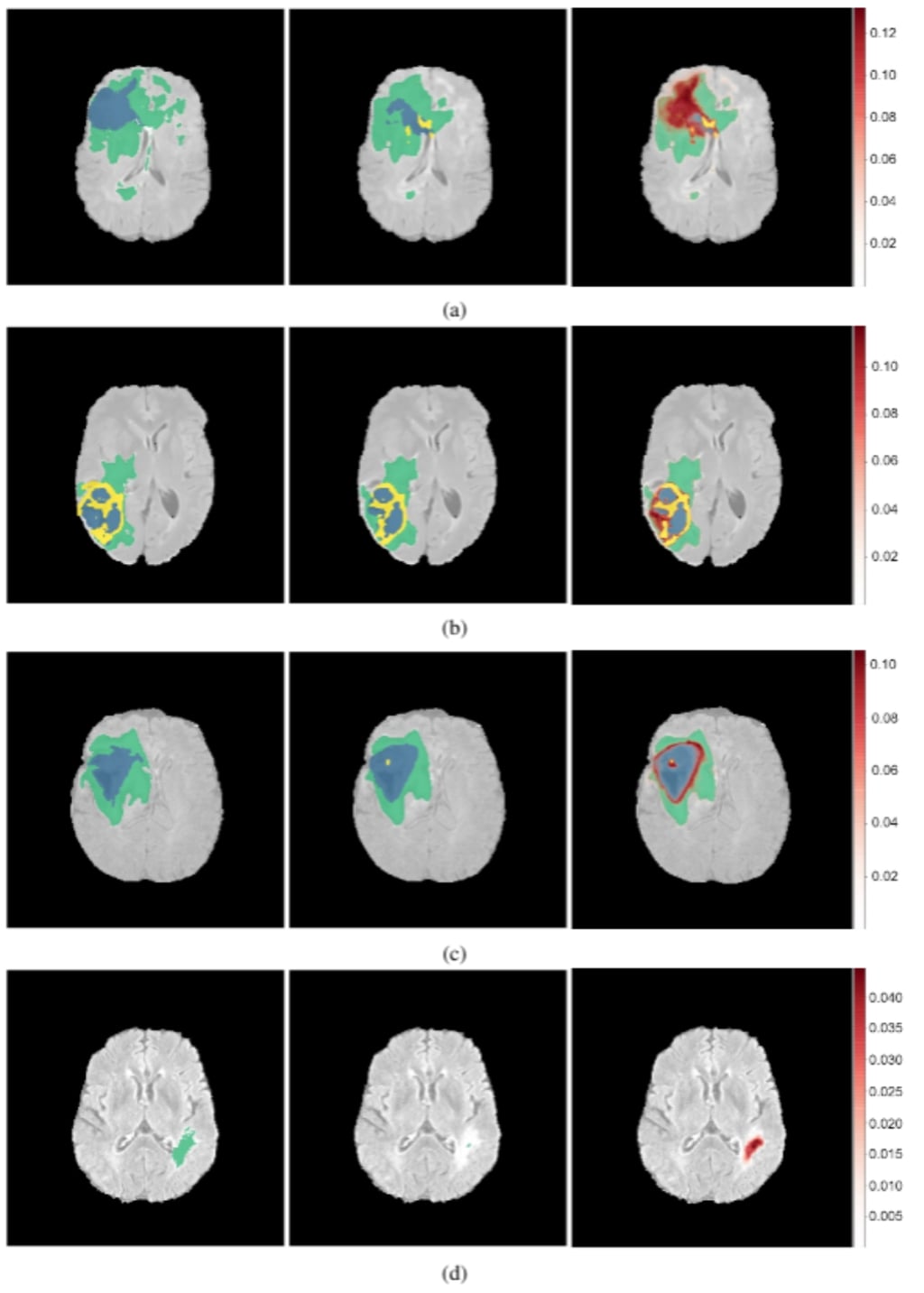}
    \caption{Uncertainty estimations (shown in red) for the DenseUnet using TTD for a selected set of images. Ground Truth(Left), Model Prediction(Middle), and Uncertainty(Right). Misclassified regions are often associated with high uncertainty.}
    \label{fig:fig5}
\end{figure}

Where $\Phi(x|{w^t})$ is the output of the neural network with weights $w^t$ on applying dropout on the $t^{th}$ iteration. The models are retrained with a dropout rate of 0.2 after each layer. At test time, a posterior distribution is generated by running the model for 100 epochs for each image. We take the mean of the posterior sampled distribution as our prediction and the channel mean of the variance from Equation \ref{eqn_unc} as the uncertainty \citep{kendall2015bayesian}. The results of this are shown in Figure \ref{fig:fig5}.

We find that regions which are misclassified are often associated with high uncertainty. For example, Figure \ref{fig:fig5}a shows a region in the upper part of the tumor which is misclassified as \textit{necrotic tumor}, but the model is also highly uncertain about this region. 
Similar behaviour is seen in Figure \ref{fig:fig5}b. In some cases, the model misses the tumor region completely, but the uncertainty map still shows that the model has low confidence in this region (\ref{fig:fig5}d), while in some cases, boundary regions are misclassified with high uncertainty (\ref{fig:fig5}c). In a medical context, these are regions that radiologists should pay more attention to. This would encourage a sort of collaborative effort - tumors are initially segmented by deep learning models and the results are then fine-tuned by human experts who concentrate only on the low-confidence regions, Figure \ref{fig:pipeline} shows.

More sample images as well as uncertainty for other networks can be found in the Supplementary Material.

% \begin{figure}
% \captionsetup[subfigure]{labelformat=empty}
% \centering
%   \subcaptionbox{(a)}{\includegraphics[width=0.9\textwidth]{unc_train_volume4.png}}\\
%   %\vspace{-4mm}
%   \subcaptionbox{(b)}{\includegraphics[width=0.9\textwidth]{unc_train_volume26.png}}\\
%   %\vspace{-4mm}
%   %\subfloat[(c)]{\includegraphics[width=0.48\textwidth]{images/uncertainty/overlay/volume18.jpg}}\hfill
%   %\subfloat[(d)]{\includegraphics[width=0.48\textwidth]{images/uncertainty/overlay/volume44.jpg}}\\
%   \subcaptionbox{(c)}{\includegraphics[width=0.9\textwidth]{unc_volume2.png}}\\
%   %\vspace{-4mm}
%   %\subfloat[(f)]{\includegraphics[width=0.48\textwidth]{images/uncertainty/overlay/volume10.jpg}}\\
%   %\subfloat[(g)]{\includegraphics[width=0.48\textwidth]{images/uncertainty/overlay/train_volume6.jpg}}\hfill
%   \subcaptionbox{(d)}{\includegraphics[width=0.9\textwidth]{unc_train_volume8.png}}
%     \caption{Uncertainty estimations (shown in red) for the DenseUnet using TTD for a selected set of images. Ground Truth(Left), Model Prediction(Middle), and Uncertainty(Right). Misclassified regions are often associated with high uncertainty.}
%     \label{fig:fig5}
% \end{figure}

% \subsection{Aleatoric Uncertainty}
% Aleatoric uncertainty captures noise inherent in the data generating process. [Gal] capture aleatoric uncertainty implicitly from the model as a learned loss attenuation term.
% [Ahyan] implement a test-time augmentation pipeline to determine aleatoric uncertainty.
% This involves augmenting the input image with minor transformations and noise.

% \begin{subequations}
% \centering
% \begin{align}
%  p(y|T(x), w) \approx \dfrac{1}{T} \sum_{t=1}^t Softmax(f^{w}(T(x)))\\
%  var_{aleatoric}(p(y|x,w)) \approx \dfrac{1}{T} \sum_{t=1}^t f^{w}(T_t(x))^T f^{w}(T_t(x)) - E(y)^TE(y)
% \end{align}
% \end{subequations}

% Where $T_t(x)$ is the transformed image at iteration t.

% We use simple spatial transforms with added gaussian noise, $T(x) = \Gamma_\beta(x) + \mathcal{N}(\mu, \sigma)$, where $\beta$ is the set of parameters of the transformation.

%\subsection{Implementation}

%We train our models with dropout after every layer. The drop probability is set at 0.2. We find three uncertainties: (1) Epistemic only with test time dropout, (2) Aleatoric only with test time augmentation, and (3) Combined uncertainty.

%For combined uncertainty, the model predictions are obtained for T iterations after applying dropout as well as augmentation at each step.

\section{Conclusion}

In this paper, we attempt to elucidate the process that neural networks take to segment brain tumors. We implement techniques for visual interpretability and concept extraction to make the functional organization of the model clearer and to extract human-understandable traces of inference.

\noindent From our introductory study, we make the following inferences:

\begin{itemize}
    \item Disentangled, human-understandable concepts are learnt by filters of brain tumor segmentation models, across architectures.
    \item Models take a largely hierarchical approach to tumor localization. In fact, the model with the best test performance shows a clear convergence from larger structures to smaller structures. 
    \item Skip and residual connections may play a role in transferring spatial information to shallower layers.
    \item Constrained optimization helps to extract feature visualizations which show distinct shapes and patterns which may be representations of tumor structures. Correlating these with the disentangled concepts extracted from Network Dissection experiments might help us understand how exactly a model detects and generalizes such concepts on a filter level.
    \item Misclassified tumor regions are often associated with high uncertainty, which indicates that an efficient pipeline which combines deep networks and fine-tuning by medical experts can be used to get accurate segmentations.
\end{itemize}

As we have discussed in the respective sections, each of these inferences might have an impact on our understanding of deep learning models in the context of brain tumor segmentation. 

While more experiments on a broader range of models and architectures would be needed to determine if such behavior is consistently seen, the emergence of such human-understandable concepts and processes might aid in the integration of such methods in medical diagnosis - a model which seems to take human-like steps is easier to trust than one that takes completely abstract and incoherent ones. This is also encouraging from a neuroscience perspective - if model behaviour is consistent with visual neuroscience research on how the human brain processes information, as some of our results indicate, this could have implications in both machine learning and neuroscience.

\section{Future Work}
\label{section7}

Future work will be centered around gaining a better understanding of the segmentation process for a greater range of models (including 3D models) and better constrained optimization techniques for extracting human-understandable feature visualizations which would allow an explicit understanding of how models learn generalized concepts. For instance, it would be worth-wile to understand what set of regularizers generates the most medically relevant images. Textural information extracted from the optimized pre-images can also be analyzed to determine their correlation with histopathological features. 

Further exploration regarding how these results are relevant from a neuroscience perspective can also be done, which might aid in understanding not just the machine learning model, but also how the brain processes information.
The inferences from our explainability pipeline can also be used to integrate medical professionals into the learning process by providing them with information about the internals of the model in a form that they can understand.

\section*{Conflict of Interest Statement}
% %All financial, commercial or other relationships that might be perceived by the academic community as representing a potential conflict of interest must be disclosed. If no such relationship exists, authors will be asked to confirm the following statement: 

The authors declare that the research was conducted in the absence of any commercial or financial relationships that could be construed as a potential conflict of interest.

\section*{Author Contributions}
 PN and AK developed the pipeline and performed the analysis and implementation. PN wrote the first draft, PN and AK revised the manuscript and generated the visualizations. GK edited the manuscript, supervised and funded the study.
% The Author Contributions section is mandatory for all articles, including articles by sole authors. If an appropriate statement is not provided on submission, a standard one will be inserted during the production process. The Author Contributions statement must describe the contributions of individual authors referred to by their initials and, in doing so, all authors agree to be accountable for the content of the work. Please see  \href{http://home.frontiersin.org/about/author-guidelines#AuthorandContributors}{here} for full authorship criteria.

% \section*{Funding}
% Financial support for study was done by Indian Institute of Technology Madras, Departmental development funds.

% \section*{Acknowledgments}
% The author would like to thank the BRATS'2018 Challenge team for making available large annotated multi-institution data sets that used in this study.

\section*{Data Availability Statement}
Publicly available data sets were used for this study. The data sets can be found at the BRATS 2018 challenge (https://www.med.upenn.edu/sbia/brats2018/data.html) (\cite{bakas2019segmentation}, \cite{bakas2020segmentation}).

\bibliographystyle{frontiersinSCNS_ENG_HUMS} 
\bibliography{frontiers}

\section{Supplementary Material}

\subsection{Network Architectures}

\begin{figure}[H]
\centering
    \includegraphics[width=1.\textwidth]{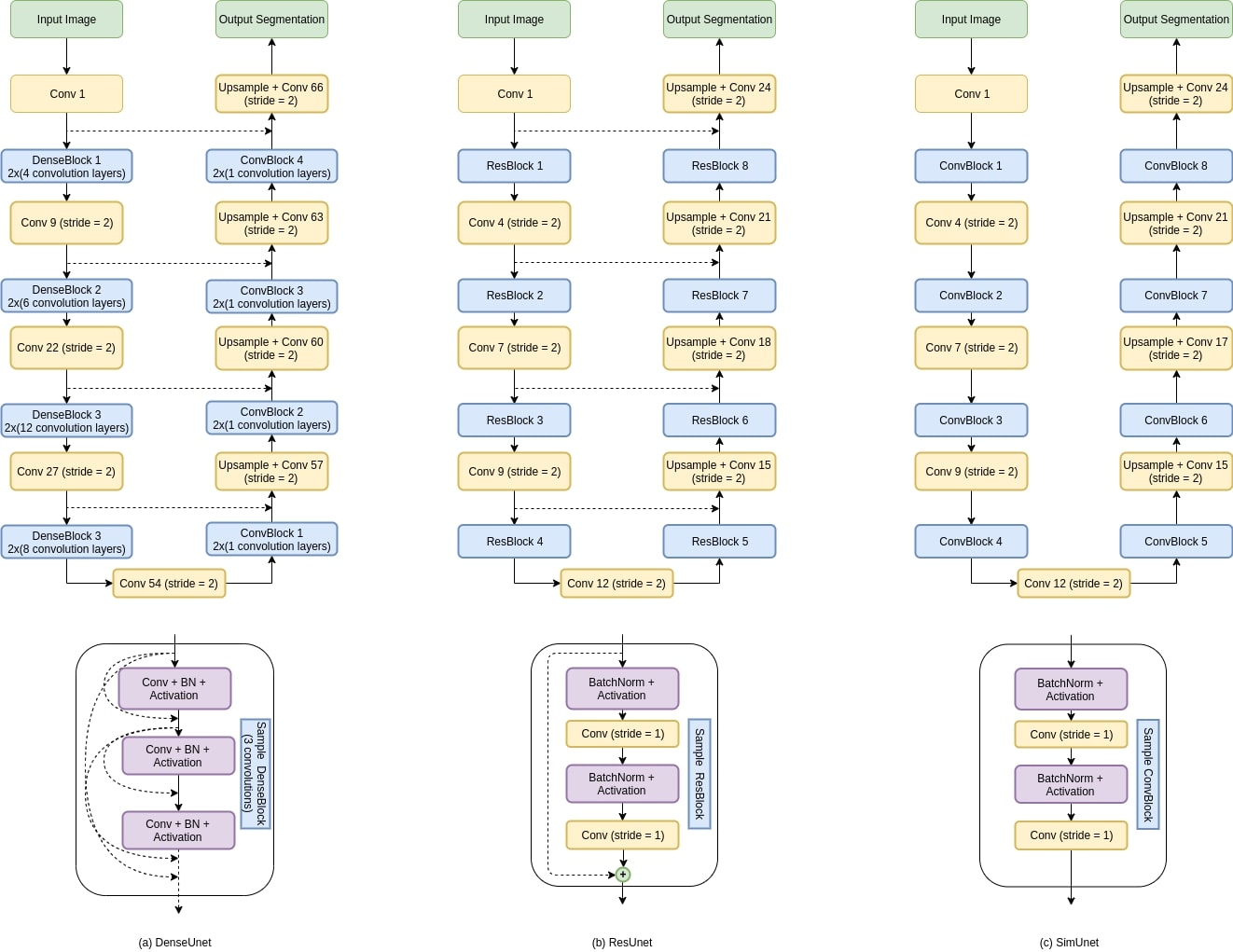}
    \caption{Network Architectures used in our study}
    \label{fig:fig8}
\end{figure}

\subsection{Network Dissection}

% \begin{figure}[H]
% \captionsetup[subfigure]{labelformat=empty}

% \centering
%   \subfloat[a]{\includegraphics[width=0.3\textwidth]{dense.png}}\hfill
%   \subfloat[b]{\includegraphics[width=0.3\textwidth]{resunet.png}}\hfill
%   \subfloat[c]{\includegraphics[width=0.3\textwidth]{simunet.png}}\\
%     \caption{Network Architectures used in our study}
%     \label{fig:fig5}
% \end{figure}

Final extracted disentangled concepts for different filters of a particular layer are shown. The figures clearly show that different filters are specialized to detect different concepts of the input image. All three networks show similar behaviour.

\begin{figure}[H]
\captionsetup[subfigure]{labelformat=empty}

    \centering
    {\includegraphics[width=0.9\textwidth]{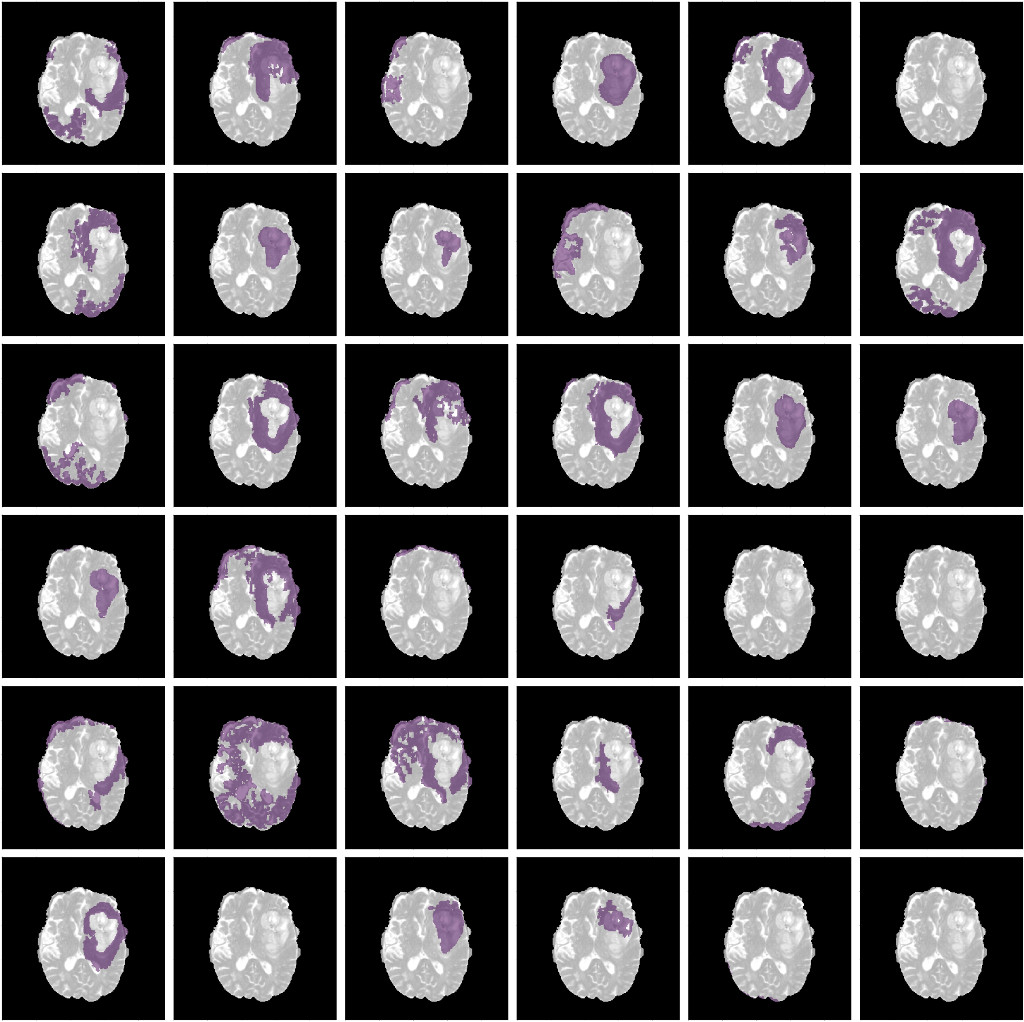}}\hfill
    \caption{Concepts learned by filters of a particular layer of the ResUnet for an input image. (Conv Layer 21)}
\end{figure}

\begin{figure}[H]
\captionsetup[subfigure]{labelformat=empty}

    \centering
    {\includegraphics[width=0.9\textwidth]{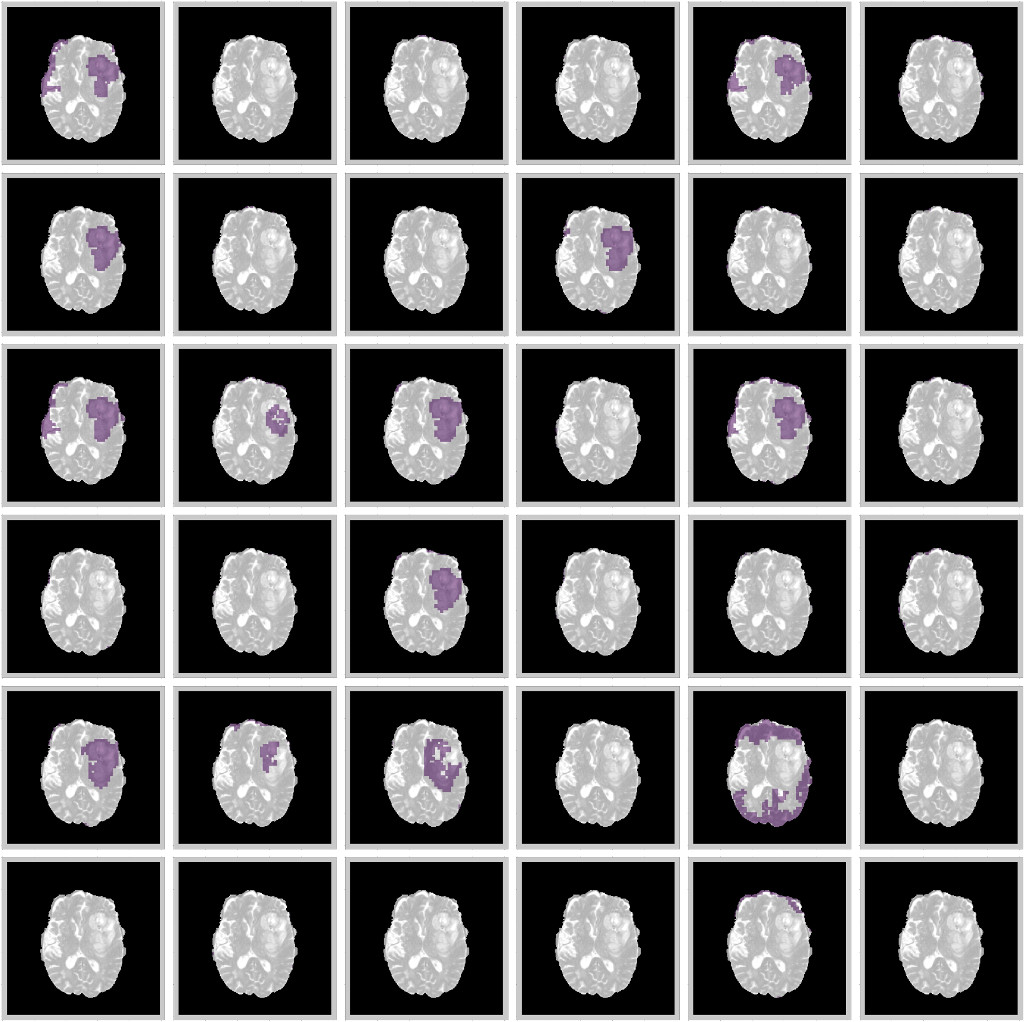}}\hfill
    \caption{Concepts learned by filters of a particular layer of the DenseUnet for an input image. (Encoding Block 1, Conv 2)}
\end{figure}

% \begin{figure}[H]
% \captionsetup[subfigure]{labelformat=empty}

%     \centering
%     {\includegraphics[width=0.9\textwidth]{sup_35_simnet_conv2d_21_.jpg}}\hfill
%     \caption{Concepts learned by filters of a particular layer of the SimUnet for an input image. (Conv Layer 21)}
% \end{figure}
\newpage

\begin{figure}[H]
\captionsetup[subfigure]{labelformat=empty}
\centering
   \subcaptionbox{(a)}{\includegraphics[width=0.4\textwidth]{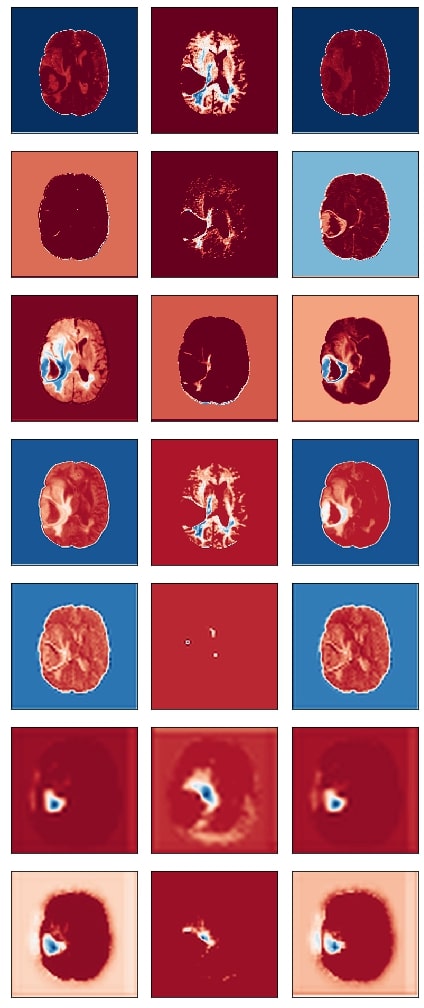}}\hfill
   \subcaptionbox{(b)}{\includegraphics[width=0.4\textwidth]{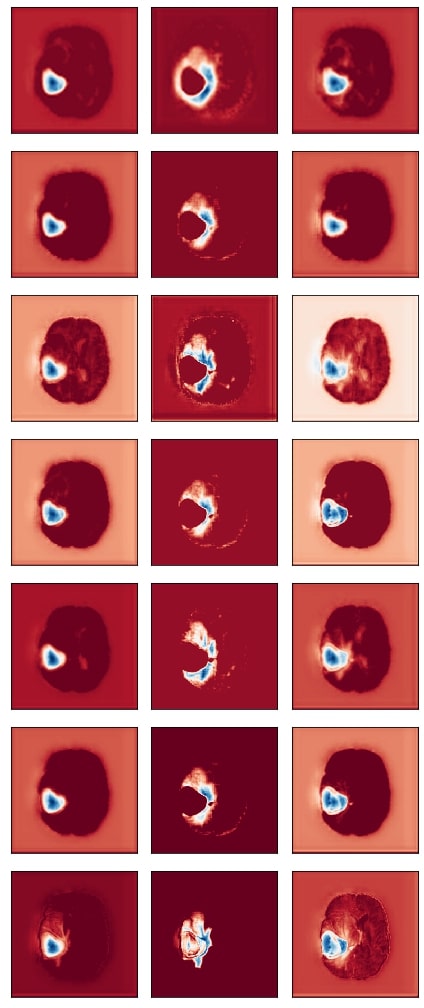}}\\
    \caption{Grad-CAM results for consecutive layers of the ResUnet (view: top to bottom, column a, followed by top to bottom, column b)}
    \label{fig:fig12}
\end{figure}

\newpage
\subsection{Feature Visualization}

The figure below shows visualized features for a randomly selected filter of successive layers. 
\begin{figure}[H]
\captionsetup[subfigure]{labelformat=empty}

    \centering
    {\includegraphics[width=0.8\textwidth]{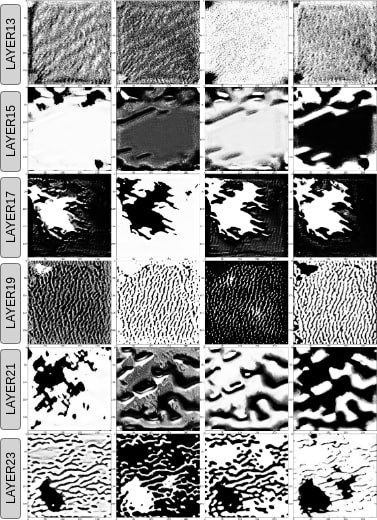}}\hfill 
    % \label{fig:fig11}
    \caption{Activation maps for layers of the ResUnet}
\end{figure}

\newpage
\begin{figure}[H]
\captionsetup[subfigure]{labelformat=empty}

    \centering
    {\includegraphics[width=0.95\textwidth]{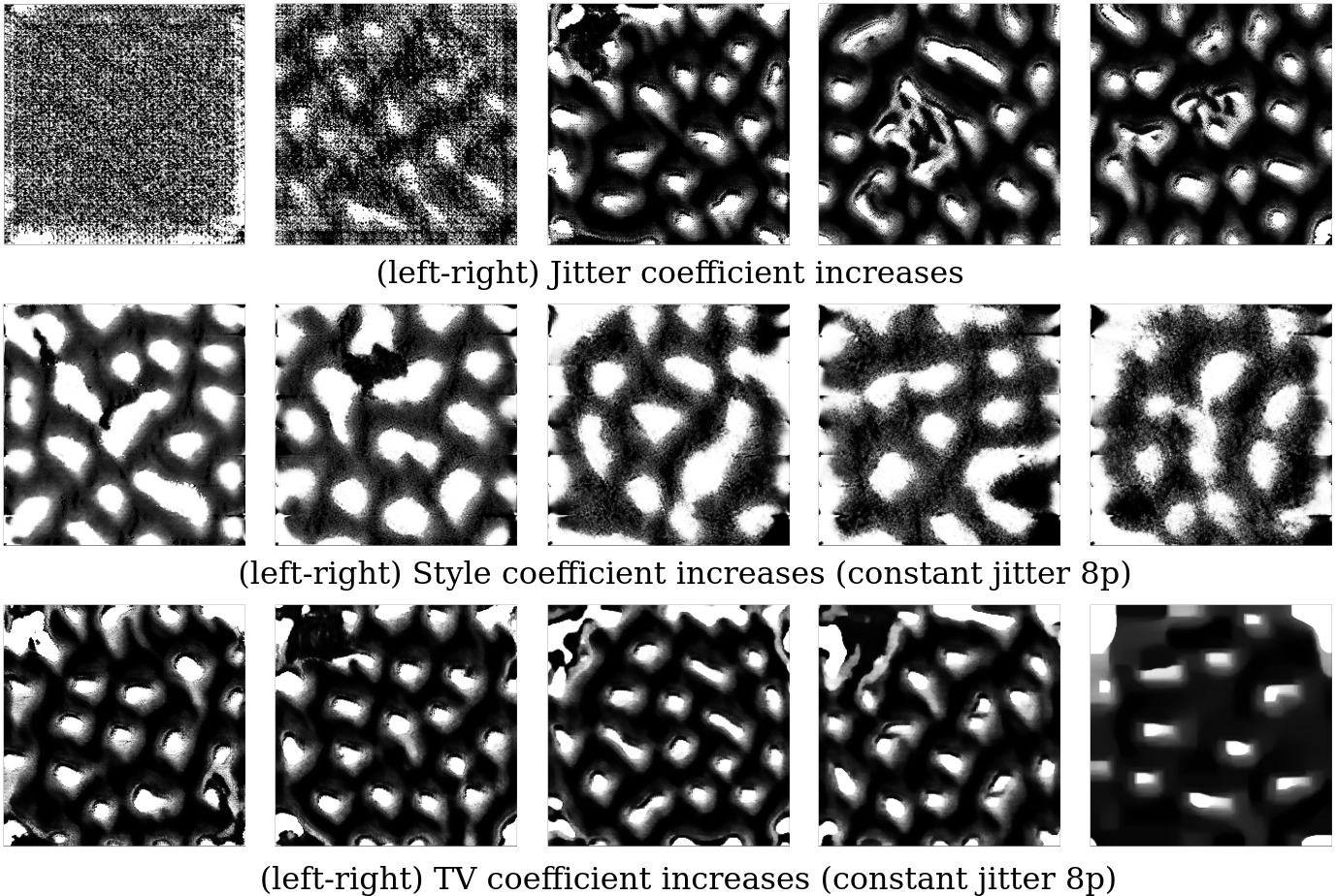}}\hfill 
    \label{fig:fig13}
    \caption{Effect of independently changing hyperparamaters for each regularizer. Top: Jitter coefficient increases [0 pixels, 1p, 6p, 12p, 20p], Middle: Style Coefficient increases [$10^{-2}$, $10^{-1}$, $1$, $5$, $10$], Bottom: Total Variation regularization increases [$10^{-7}$, $10^{-6}$, $10^{-5}$, $10^{-4}$, $10^{-3}$] to smoothen image}
\end{figure}

\subsection{Uncertainty}

\begin{figure}[H]
\captionsetup[subfigure]{labelformat=empty}
\centering
   {\includegraphics[width=0.9\textwidth]{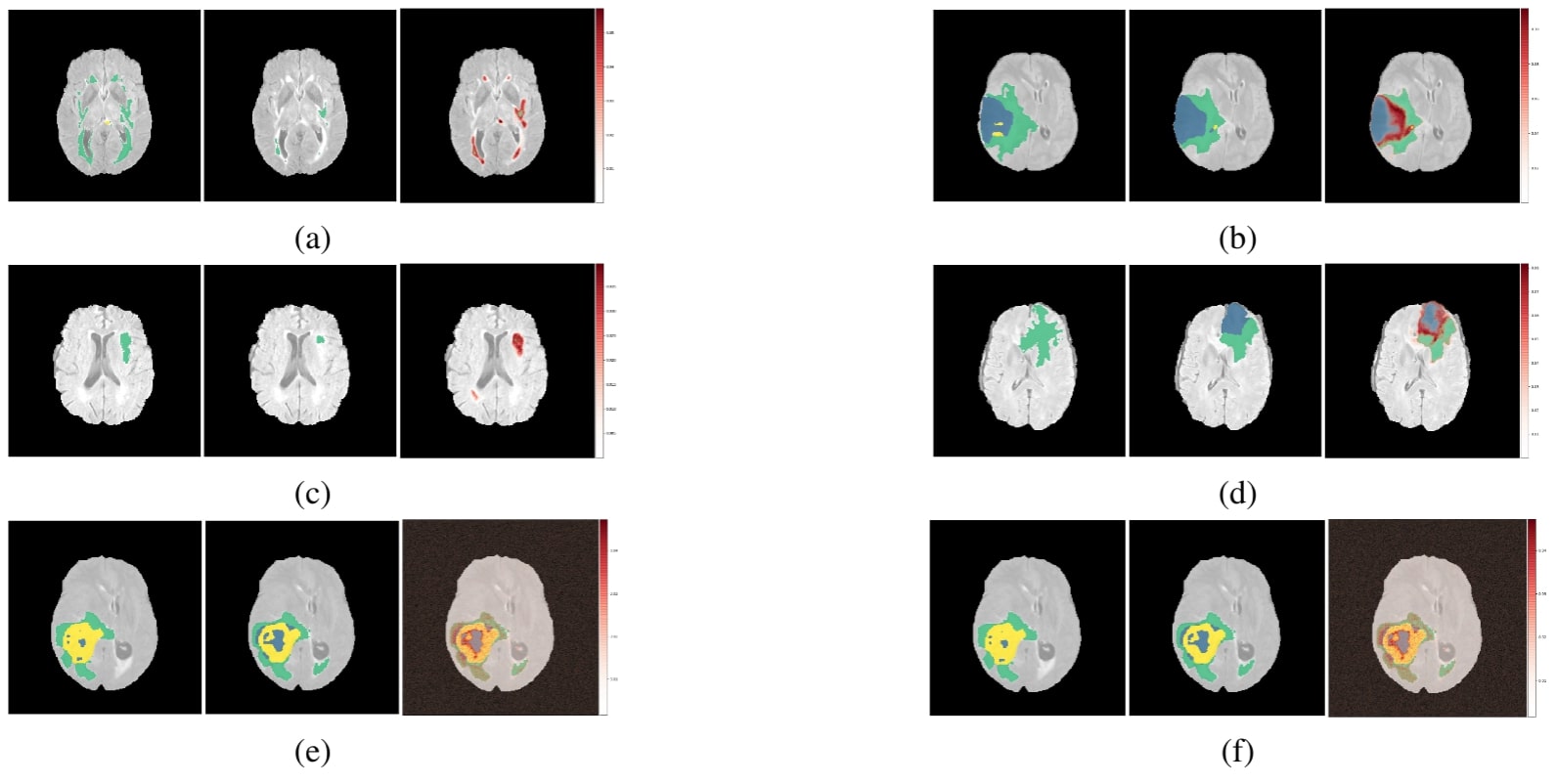}}\hfill
    \caption{Uncertainty estimations (shown in red) for the DenseUnet (a,b,c,d) and ResUnet (e,f). Ground Truth(Left), Model Prediction(Middle), and Uncertainty(Right).}
    \label{fig:fig14}
\end{figure}

\end{document}